# Low state transitions in the nova-like cataclysmic variable HS 0506+7725

Jeremy Shears

## Abstract

The twelve-year light curve of the nova-like cataclysmic variable HS 0506+7725 between 2006 April and 2018 November is presented. This shows that the star spends most of the time in a normal state at magnitude ~14.7, but multiple excursions to a fainter state at magnitude 16.0 to 17.0 were apparent. These normal state/low state transitions of up to 2.3 magnitudes are typical of the VY Scl subclass of CVs. The second of these fading episodes was the best characterised as its return to normal brightness was also observed. The complete transition lasted about 43 days. Further monitoring of this system by amateur astronomers is encouraged to identify and characterise future low states.

## Introduction

Cataclysmic variables (CVs) are semi-detached binary stars in which a white dwarf primary accretes material from a secondary star via Roche lobe overflow. The secondary is usually a late-type main sequence star. In the absence of a significant white dwarf magnetic field, material from the secondary is processed through an accretion disc before settling on the surface of the white dwarf. In CVs with high mass transfer rates the accretion disc becomes thermally stable. Such systems are called nova-like stars, because they lack outbursts characteristic of dwarf novae (1). Among nova-like stars, there exists a small group known as VY Scl stars which exhibit pronounced low states in which the system fades by up to 5 magnitudes at irregular intervals. These low states are due to the temporary reduction or cessation of mass transfer from the secondary star to the white dwarf primary (2). It is thought that the accretion flow is disrupted when a star spot on the secondary drifts across the point where the accretion stream leaves the secondary (3).

HS 0506+7725 was identified as a nova-like CV star during follow-up observations of optically selected CV candidates from the Hamburg Quasar Survey (HQS) by Aungwerojwit *et al*. (4) (5). Its orbital period of 212.7 ± 0.2 min (0.14771 d, 3.5450 h) was determined from radial velocity and photometric variability studies.

HS 0506+7725 was characterised by strong Balmer and He emission lines, short-period (~10–20 min) optical flickering, and weak X-ray emission in the ROSAT All Sky Survey. The relatively narrow emission lines and the low amplitude of the radial velocity suggested a low inclination. The detection of a deep low state (B~ 18.3) identified HS 0506+7725 as a member of the VY Scl subclass

In 2006, Prof. Boris Gänsicke (University of Warwick), whose team had characterised many of the HQS CVs including HS 0506+7725, suggested that the author might like to carry out a long-term photometric monitoring of the star. The results of the 12-year study are presented here.

HS 0506+7725 is located at RA 05 13 36.57, Dec. +77 28 42.8 (J2000.0) in Camelopardalis. It is also catalogued in the first ROSAT X-ray Survey as 1RXS J051336.1+772836.

## Observations

The author performed unfiltered CCD photometry on HS 0506+7725 using V-band comparison stars from AAVSO sequence X23683LR (6). Before 2007 Sep a 0.1 m refractor was used with a Starlight Xpress MX716 CCD camera and after this time a 0.28 m SCT was deployed with a Starlight Xpress SXV-H9 camera. V-band photometry from the All Sky Automated Survey for Supernovae (ASAS-SN) was also used, covering the interval 2013 Aug 1 and 2018 Nov 12. This is an automated sky survey to search for new supernovae and other astronomical transients (7) (8) (9) which is able to detect and perform photometry on objects down to 18th magnitude. V-band data from the AAVSO Photometric All-Sky Survey (APASS) were also used (10).

**Results**

The long-term light curve of HS 0506+7725 between 2006 April 22 and 2018 Nov 18 is presented in Figure 1. This shows that in its normal state, the star varies between magnitude ~14.6 and 14.9; night to night variations of up to 0.4 magnitude are typical. A number of excursions to a faint state at magnitude 16.0 to 17.0 are apparent, which we interpret as VY Scl-type low states. Taking the typical normal brightness as magnitude 14.7, these low states represent fades of up to 2.3 magnitudes.

To characterise the low states further, we determined when the star faded to magnitude 16.0 or fainter, taking this as indicative of a transition to a low state. These excursions are more readily apparent in the higher resolution plots in Figure 2. This simple analysis identifies sixteen times when the star faded to magnitude 16.0 or fainter and these are listed in Table 1. It should be noted that this analysis is only partially satisfactory as the cadence of the data in the earlier years, especially before the ASAS-SN data, is rather low, therefore most of the transitions are poorly defined. It is evident that some of these excursions lasted only a few days and it is therefore likely that others might have been missed. There may be some longer ones, for example event 15 discussed below, but because of the significant gaps in the data, their duration cannot be certain. Because of this under sampling, it is also possible that not all of the 16 fading events represent separate episodes; as noted above, the star shows significant variations in its normal state and if these were to continue during a fade they might appear as a sudden brightening and fade, superimposed on a global fading trend, thus appearing as two separate events. For example, low states 1 and 2 might be part of a single event.

Two of the fading events are better characterised by the denser coverage provided by ASAS-SN: number 15 and 16. Event 15, starting around JD 2457100, resulted in the star fading at an average of 0.037 mag/d (linear fit to the data) over the next 40 days. Unfortunately, the return to the normal state was not observed, so it is not possible to say how long the low state lasted. There was a gap of 89 days after the faintest part of the fade and before the next observation which is itself evidently part of a brightening from a fade. Of course, this fading arm and brightening arm might be part of two separate events, but if they were indeed part of the same event, the total duration of event 15 would have been about 160 days.

The second better characterised event (number 16) showed both the fading arm and the brightening arm. It began at around JD 2457835 and over the next 30 days the star faded by an average of 0.034 mag/d. No sooner had the star reached a minimum of magnitude 16.1 than it began to brighten again. The return to the normal state was much more rapid taking only 13 days, with the average brightening rate being – 0.09 mag/d. Thus, the overall fading episode lasted ~43 days.

**Discussion**

The results presented here clearly show several normal state/low state transitions in HS 0506+7725 of up to 2.3 magnitudes which are typical of VY Scl nova-like CVs. Honeycutt and Kafka (2) found that the transitions to (and from) faint states exhibit some systematic properties. Most of these transitions were adequately fit by a single straight line to the magnitude light curve. They found that the average *e*-folding time, τ, was about 20 days, where τ= ((log10 *e*)/0.4)/(Δmag/Δt), although there was a considerable variation with τ = 3 to 94 days. The two fades characterised in HS 0506+7725 of 0.037 and 0.034 mag/d (event 15 and 16) correspond to τ = 29 and 32 days are therefore consistent with these values. The faster rise in the case of the event 16 of – 0.09 mag/d (τ = 12 days) is also typical of VY Scl systems as reported by Honeycutt and Kafka (2).

It is noteworthy that much of the long-term photometry reported in this paper was obtained with small amateur telescopes, which highlights the continuing value of amateur astronomers' observations even in the era of wide-field transient surveys. In any case, the high northerly declination of this object puts it out of range of some surveys. It is gratifying that ASAS-SN now reports photometry on this object, but lengthy gaps in the survey data still exist. It is therefore important to continue monitoring HS 0506+7725 to determine how frequently it undergoes low states and to what extent these states vary in duration and brightness. It is surprising that apart from the author, no other amateur astronomer appears to have been observing this interesting object, according to the AAVSO International Database. A consequence of the lack of observational data is that the frequency and duration of these low states is very poorly characterised.

There is potentially great value in identifying future low states of HS 0506+7725 since at these times accretion in nova-like systems may reduce to a low level or sometimes switch off completely. At such times observations with professional-class instrumentation might reveal the white dwarf and spectroscopy might shed further light on the nature of the secondary star which can be observed without its light being swamped by that of the accretion disc. Boris Gänsicke has asked to be notified of future low states.

The light curve presented here confirms Aungwerojwit *et al.*'s (5) conclusion, based on a single low state observation, that HS 0506+7725 is a VY Scl system. However, the brightness they reported, B~18.3, is much fainter than any low state reported here, although this was using a B-filter so cannot be directly compared with the data reported in this paper. It is possible that this was indeed a true low state with minimal or no accretion. It is hoped that such an event will be observed again in the future.

**Conclusions**

The twelve-year light curve of HS 0506+7725 shows that the star spends most of the time in a normal state at magnitude 14.6 to 14.9, but a number of excursions to a faint state at magnitude 16.0 to 17.0 were apparent. Such normal state/low state transitions of up to 2.3 magnitudes are typical of the VY Scl nova-like CVs. Moreover, the rate of fading during two well-observed episodes had *e*-folding times of τ = 29 and 32 days which is also consistent with other VY Scl systems. The second of these fading episodes was the best characterised and lasted about 43 days; the subsequent brightening was much more rapid than the fade (τ = 12 days).

HS 0506+7725 is an under observed CV and it is hoped that other amateur astronomers will add it to their observational programmes in the hope of observing and characterising future low states.

**Acknowledgments**

The author is grateful to Professor Boris Gänsicke, University of Warwick, for encouraging him to take on this project, but also for sharing his enthusiasm and knowledge over many years.

The research made use of data from the All Sky Automated Survey for Supernovae, AAVSO Photometric All-Sky Survey, the AAVSO Variable Star Index, the NASA/Smithsonian Astrophysics Data System and SIMBAD, operated through the Centre de Données Astronomiques (Strasbourg, France).

*JS: "Pemberton", School Lane, Bunbury, Tarporley, Cheshire, CW6 9NR, UK*
*[bunburyobservatory@hotmail.com]*

**References**

**1.** *Warner, B. Cataclysmic Variable Stars (Cambridge University Press) 1995.*

**2.** *Honeycutt R.K. and Kafka S., AJ, 128, 1279-1293 (2004).*

**3.** *Livio M., & Pringle J. E., ApJ, 427, 956-960 (1994).*

**4.** *Aungwerojwit A., Gänsicke B.T., Rodríguez-Gil P. et al., ASP Conference Series, 330, 469-470 (2005).*

**5.** *Aungwerojwit A., Gänsicke B.T., Rodríguez-Gil P. et al., A&A, 443, 995-1005 (2005).*

**6.** *Charts and sequence information can be obtained from the AAVSO Variable Star Plotter at https://www.aavso.org/apps/vsp/.*

**7.** *Information about the ASAS-SN is available at https://asas-sn.osu.edu/.*

**8.** *Shappee B.J. et al., ApJ, 788, article id. 48 (2014).*

**9.** *Kochanek C.S. et al., PASP, 129, 104502 (2017).*

**10.** *https://www.aavso.org/apass.*

| Event number | JD | Date | Magnitude at minimum | Depth |
|---|---|---|---|---|
| 1 | 2454024 | 2006 Oct 15 | 16.4 | 1.7 |
| 2 | 2454048 | 2006 Nov 8 | 16.4 | 1.7 |
| 3 | 2454116 | 2007 Jan 15 | 16.1 | 1.4 |
| 4 | 2454160 | 2007 Feb 28 | 16.0 | 1.3 |
| 5 | 2454835 | 2009 Jan 3 | 16.0 | 1.3 |
| 6 | 2455078 | 2009 Sep 3 | 16.8 | 2.1 |
| 7 | 2455152 | 2009 Nov 16 | 17.0 | 2.3 |
| 8 | 2455233 | 2010 Feb 5 | 16.3 | 1.6 |
| 9 | 2455291 | 2010 Apr 4 | 16.4 | 1.7 |
| 10 | 2455333 | 2010 May 16 | 16.8 | 2.1 |
| 11 | 2455443 | 2010 Sep 3 | 16.1 | 1.4 |
| 12 | 2456272 | 2012 Dec 10 | 16.2 | 1.5 |
| 13 | 2456308 | 2013 Jan 15 | 16.3 | 1.6 |
| 14 | 2456528 | 2013 Aug 23 | 16.1 | 1.4 |
| 15 | 2457127 | 2015 Apr 14 | 16.3 | 1.6 |
| 16 | 2457856 | 2017 Apr 12 | 16.1 | 1.4 |

**Table 1: Low states in HS 0506+7725**

For each event, the date at which the star first reached magnitude 16.0 is given, along with the faintest magnitude it reached during that low state and the depth compared to an arbitrary normal state magnitude of 14.7

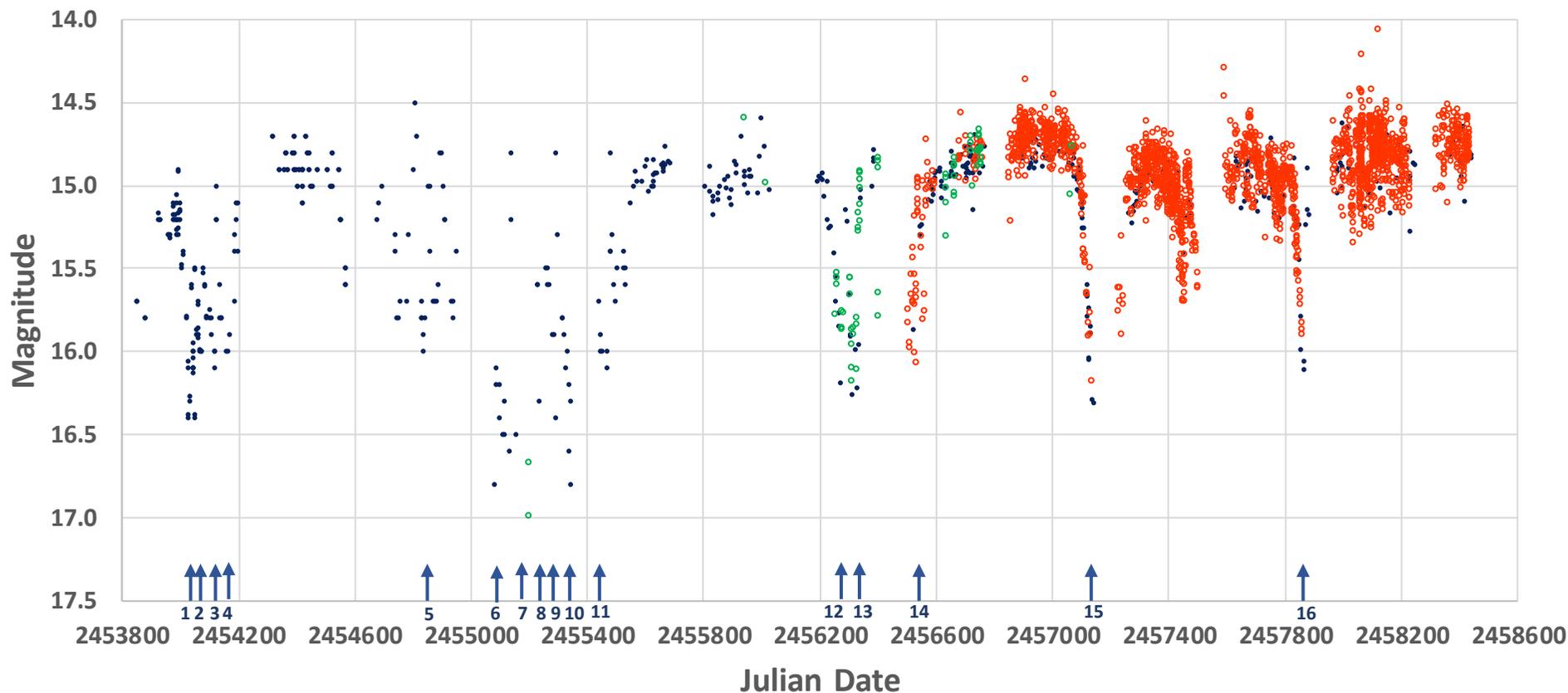

Figure 1: Light curve of HS 0506+7725 between 2006 April 22 and 2018 Nov 18

Blue data: CV photometry by the author. Red data: ASAS-SN V-band. Green data: AAVSO Photometric All-Sky Survey (APASS) V-band data. The blue arrows show when the star faded to 16.0 or fainter

*Next page*

Figure 2: Combined data from Figure 1 replotted at a larger scale

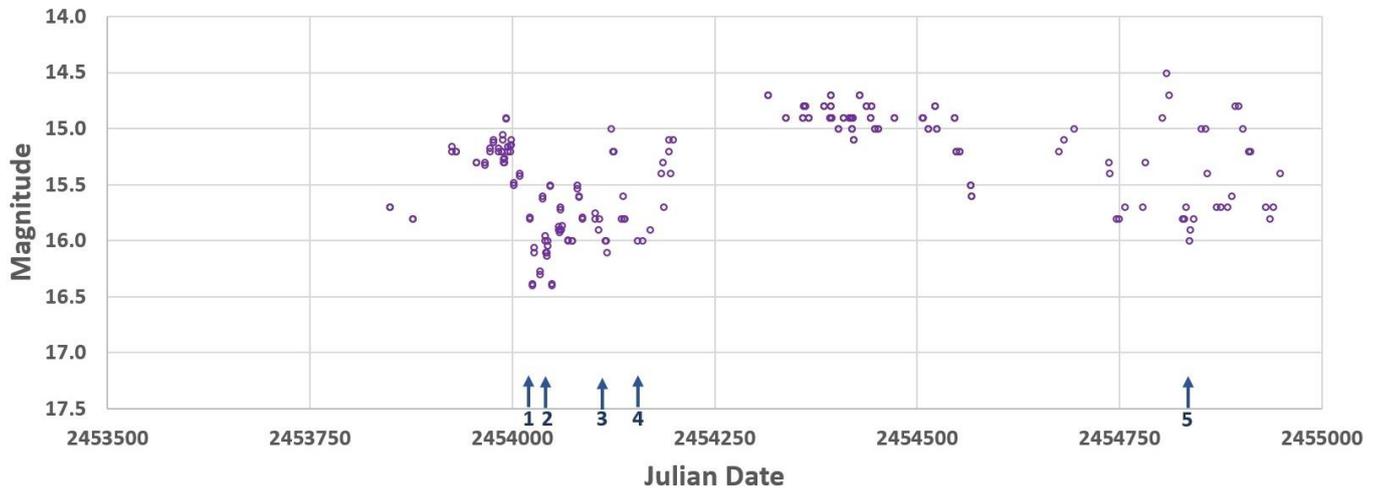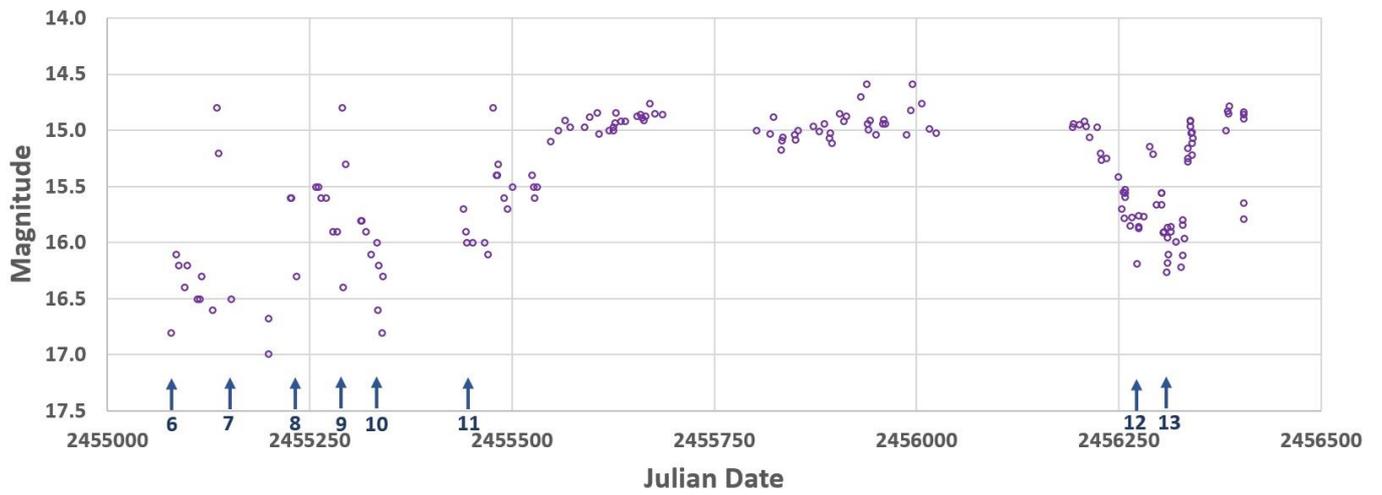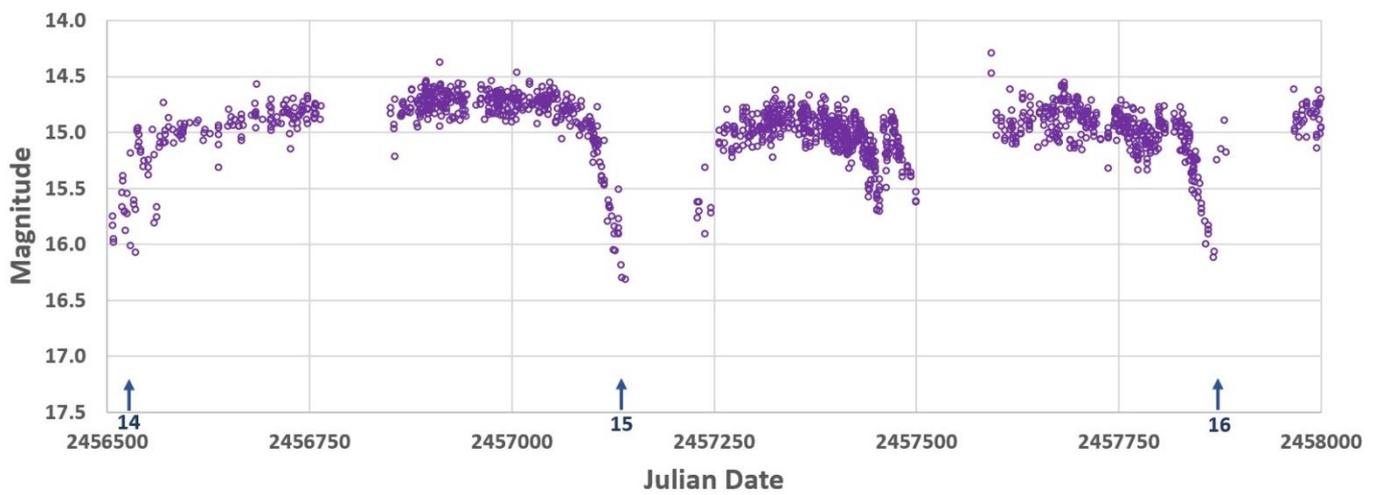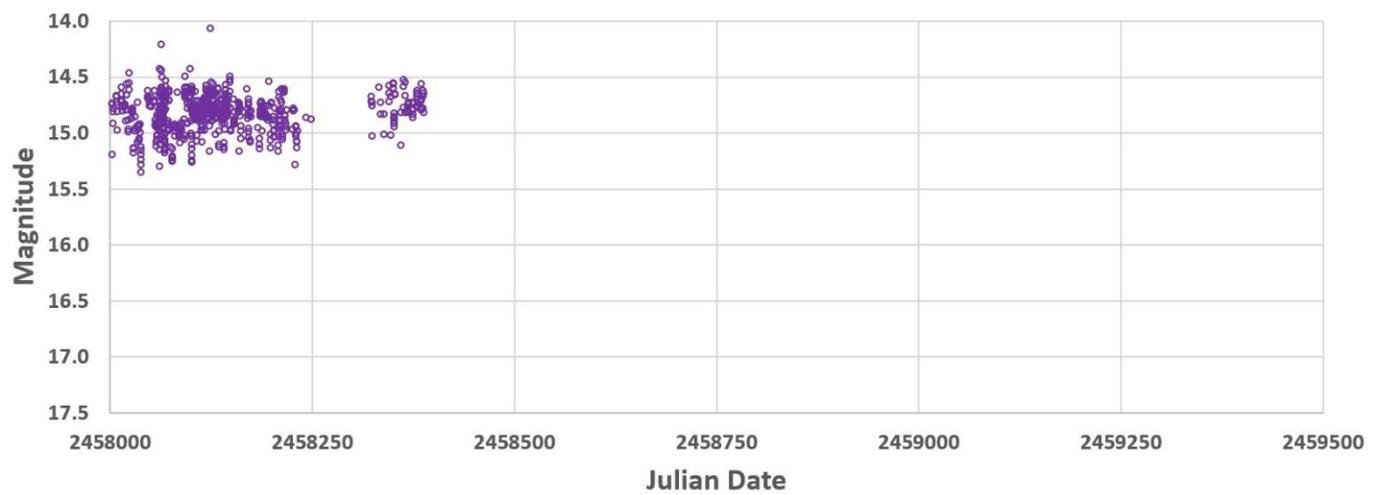